\documentclass[aps,prd, twocolumn,superscriptaddress]{revtex4-1}

\usepackage{graphicx}
\usepackage{dcolumn}
\usepackage{bm}
\usepackage{amssymb}   
\usepackage{aas_macros}
\usepackage[hyperindex,breaklinks]{hyperref}

\newcommand{\kcmb}{\kappa_\mathrm{CMB}}
\newcommand{\kgal}{\kappa_\mathrm{gal}}
\newcommand{\arcsec}{^{\prime\prime}}

\begin{document}

\title{First Measurement of the Cross-Correlation of CMB Lensing and Galaxy Lensing}


\author{Nick~Hand} \email{nhand@berkeley.edu}
\affiliation{Dept. of Astronomy, University of California, Berkeley, CA, USA 94720}

\author{Alexie~Leauthaud}
\affiliation{Kavli Institute for the Physics and Mathematics of the Universe (WPI), 
Todai Institutes for Advanced Study, the University of Tokyo, Kashiwa, Japan}

\author{Sudeep~Das}
\affiliation{High Energy Physics Division, Argonne National Laboratory, 9700 S Cass Avenue, Lemont, IL 60439}
\affiliation{Berkeley Center for Cosmological Physics, Berkeley, CA, USA 94720}

\author{Blake~D.~Sherwin}
\affiliation{Dept. of Physics, University of California, Berkeley, CA, USA 94720}
\affiliation{Miller Institute for Basic Research in Science, University of California, Berkeley, CA, USA 94720}
\affiliation{Berkeley Center for Cosmological Physics, Berkeley, CA, USA 94720}

\author{Graeme~E.~Addison}
\affiliation{Dept. of Physics and Astronomy, University of British Columbia, 
6224 Agricultural Road, Vancouver, V6T 1Z1, BC, Canada}

\author{J.~Richard~Bond}
\affiliation{Canadian Institute for Theoretical Astrophysics, University of
Toronto, Toronto, ON, Canada M5S 3H8}

\author{Erminia~Calabrese}
\affiliation{Sub-department of Astrophysics, University of Oxford, Keble Road, Oxford OX1 3RH, UK}

\author{Ald\'ee~Charbonnier}
\affiliation{Observat\'orio do Valongo, Universidade Federal do Rio de Janeiro, 
Ladeira do Pedro Ant\^onio 43, Sa\'ude, Rio de Janeiro, RJ 20080-090, Brazil}
\affiliation{Centro Brasileiro de Pesquisas F\'{i}sicas, Rua Dr. Xavier Sigaud 150, 
CEP 22290-180, Rio de Janeiro, RJ, Brazil}

\author{Mark~J.~Devlin}
\affiliation{Dept. of Physics and Astronomy, University of
Pennsylvania, 209 South 33rd Street, Philadelphia, PA, USA 19104}

\author{Joanna~Dunkley}
\affiliation{Sub-department of Astrophysics, University of Oxford, Keble Road, Oxford OX1 3RH, UK}

\author{Thomas~Erben}
\affiliation{Argelander Institute for Astronomy, University of Bonn, 
Auf dem H\"ugel 71, 53121 Bonn, Germany}

\author{Amir~Hajian}
\affiliation{Canadian Institute for Theoretical Astrophysics, University of
Toronto, Toronto, ON, Canada M5S 3H8}

\author{Mark~Halpern}
\affiliation{Dept. of Physics and Astronomy, University of British Columbia, 
6224 Agricultural Road, Vancouver, V6T 1Z1, BC, Canada}

\author{Joachim~Harnois-D\'eraps}
\affiliation{Dept. of Physics and Astronomy, University of British Columbia, 
6224 Agricultural Road, Vancouver, V6T 1Z1, BC, Canada}
\affiliation{Canadian Institute for Theoretical Astrophysics, University of
Toronto, Toronto, ON, Canada M5S 3H8}
\affiliation{Dept. of Physics, University of Toronto, 
60 St. George Street, Toronto, ON, Canada M5S 1A7}

\author{Catherine~Heymans}
\affiliation{Scottish Universities Physics Alliance, Institute for Astronomy, 
University of Edinburgh, Royal Observatory, Blackford Hill, Edinburgh, EH9 3HJ, UK}

\author{Hendrik~Hildebrandt}
\affiliation{Argelander Institute for Astronomy, University of Bonn, 
Auf dem H\"ugel 71, 53121 Bonn, Germany}

\author{Adam~D.~Hincks}
\affiliation{Dept. of Physics and Astronomy, University of British Columbia, 
6224 Agricultural Road, Vancouver, V6T 1Z1, BC, Canada}

\author{Jean-Paul~Kneib}
\affiliation{Laboratoire d'Astrophysique (LASTRO), Ecole Polytechnique F\'ed\'erale de Lausanne 
(EPFL), Observatoire de Sauverny, CH-1290 Versoix, Switzerland}
\affiliation{Aix Marseille Universit\'e, CNRS, LAM 
(Laboratoire d'Astrophysique de Marseille) UMR 7326, 13388, Marseille, France}

\author{Arthur~Kosowsky}
\affiliation{Dept. of Physics and Astronomy, University of Pittsburgh, 
Pittsburgh, PA, USA 15260}

\author{Martin~Makler}
\affiliation{Centro Brasileiro de Pesquisas F\'{i}sicas, Rua Dr. Xavier Sigaud 150, 
CEP 22290-180, Rio de Janeiro, RJ, Brazil}

\author{Lance~Miller}
\affiliation{Dept. of Physics, University of Oxford, Keble Road, Oxford OX1 3RH, UK}

\author{Kavilan~Moodley}
\affiliation{Astrophysics and Cosmology Research Unit, School of
Mathematical Sciences, University of KwaZulu-Natal, Durban, 4041,
South Africa}

\author{Bruno~Moraes}
\affiliation{Centro Brasileiro de Pesquisas F\'{i}sicas, Rua Dr. Xavier Sigaud 150, 
CEP 22290-180, Rio de Janeiro, RJ, Brazil}
\affiliation{Department of Physics and Astronomy, University College London, Gower Street, London, WC1E 6BT, UK}
\affiliation{CAPES Foundation, Ministry of Education of Brazil, Brasilia/DF 70040-020, Brazil}

\author{Michael~D.~Niemack}
\affiliation{Dept. of Physics, Cornell University, Ithaca, NY 14853}

\author{Lyman~A.~Page}
\affiliation{Dept. of Physics, Princeton University, Princeton, NJ, USA 08544}

\author{Bruce~Partridge}
\affiliation{Dept. of Physics and Astronomy, Haverford College,
Haverford, PA, USA 19041}

\author{Neelima~Sehgal}
\affiliation{Dept. of Physics and Astronomy, 
Stony Brook University, Stony Brook, NY 11794-3800, USA}

\author{Huanyuan~Shan}
\affiliation{Laboratoire d'Astrophysique (LASTRO), Ecole Polytechnique F\'ed\'erale de Lausanne 
(EPFL), Observatoire de Sauverny, CH-1290 Versoix, Switzerland}

\author{Jonathan~L.~Sievers}
\affiliation{Astrophysics and Cosmology Research Unit, School of
Mathematical Sciences, University of KwaZulu-Natal, Durban, 4041,
South Africa}
\affiliation{Dept. of Physics, Princeton University, Princeton, NJ, USA 08544}
\affiliation{Canadian Institute for Theoretical Astrophysics, University of
Toronto, Toronto, ON, Canada M5S 3H8}

\author{David~N.~Spergel}
\affiliation{Dept. of Astrophysical Sciences, Peyton Hall, 
Princeton University, Princeton, NJ USA 08544}

\author{Suzanne~T.~Staggs}
\affiliation{Dept. of Physics, Princeton University, Princeton, NJ, USA 08544}

\author{Eric~R.~Switzer}
\affiliation{NASA/Goddard Space Flight Center, Greenbelt, MD, USA 20771}
\affiliation{Canadian Institute for Theoretical Astrophysics, University of
Toronto, Toronto, ON, Canada M5S 3H8}

\author{James~E.~Taylor}
\affiliation{Dept. of Physics and Astronomy, University of Waterloo, 
Waterloo, Ontario, Canada N2L 3G1}

\author{Ludovic~Van~Waerbeke}
\affiliation{Dept. of Physics and Astronomy, University of British Columbia, 
6224 Agricultural Road, Vancouver, V6T 1Z1, BC, Canada}

\author{Charlotte~Welker}
\affiliation{Institut d'Astrophysique de Paris, 98 bis boulevard Arago, 75014 Paris, France}
\affiliation{UPMC Paris VI, 4 place Jussieu 75005 Paris, France}

\author{Edward~J.~Wollack}
\affiliation{NASA/Goddard Space Flight Center, Greenbelt, MD, USA 20771}


\date{\today}

\begin{abstract}
We measure the cross-correlation of cosmic microwave background lensing convergence 
maps derived from Atacama Cosmology Telescope data with galaxy lensing convergence 
maps as measured by the Canada-France-Hawaii Telescope Stripe 82 Survey. The CMB-galaxy lensing cross
power spectrum is measured for the first time with a significance of $4.2\sigma$, 
which corresponds to a 12\% constraint on the amplitude of density fluctuations at redshifts $\sim$ 0.9. 
 With upcoming improved lensing data, this novel type of measurement will become a 
 powerful cosmological probe, providing a precise measurement of the mass distribution 
 at intermediate redshifts and serving as a calibrator for systematic biases in 
 weak lensing measurements.
\end{abstract}


\maketitle

\section{Introduction}
The cosmic web of matter gravitationally deflects the paths of photons
as they traverse the Universe -- an effect known as gravitational lensing. In the case of light from 
the cosmic microwave background (CMB), 
these lensing deflections imprint information about the density fluctuations between the 
primordial Universe at $z \sim$ 1100 and the present day onto the observed CMB sky, 
and in doing so modify the statistical
 properties of the CMB anisotropies.
Similarly, cosmological information about the lower-redshift Universe
can be extracted from the lensing-induced distortion of the shapes of galaxies, an effect referred
to as weak lensing. In both cases, precise measurements of the small magnification and shear 
effects can be used to reconstruct the convergence field, which is a direct measure 
of the projected matter density \cite{Lewis:2006aa, Munshi:2008aa}.

Previous analyses have demonstrated the sensitivity of CMB lensing to the 
large-scale dark matter distribution through cross-correlations with
sources that trace the same structure in the low-redshift Universe. To date, 
several galaxy catalogs, the cosmic infrared background, and quasars have been 
shown to be well-correlated with the CMB lensing convergence field 
\cite{Feng:2012aa, Bleem:2012aa, Holder:2013aa, Planck-Collaboration:2014ab, Sherwin:2012aa, Geach:2013aa}. 
Here, we report the first cross-correlation between CMB lensing and galaxy lensing through a 
measurement of the lensing-lensing cross power spectrum. The detection is a direct 
measure of the mass distribution localized to intermediate redshifts solely through 
the gravitational effects of lensing. It is also nearly insensitive to residual 
systematics that are independent in both data sets, providing a robust test 
of the $\Lambda$CDM model on the largest cosmic scales.

Lensing measurements are sensitive to both the expansion and growth histories of the Universe 
\cite{Hu:1999aa, de-Putter:2009aa, Das:2012aa}. Separately, measurements of CMB lensing 
\cite{Das:2014aa, Sherwin:2011aa, van-Engelen:2012aa, Planck-Collaboration:2014aa} 
and galaxy lensing \cite{Benjamin:2013aa, Heymans:2013aa, Kilbinger:2013aa}
have already contributed to strong constraints on the amplitude of matter fluctuations
and the nature of dark energy. Correlating weak 
lensing effects on the CMB and galaxies can break previous parameter degeneracies and offer 
powerful constraints on the evolution and nature of dark energy, the amplitude
of matter fluctuations, and the sum of neutrino masses 
\cite{Hu:2002aa, Hollenstein:2009aa, Namikawa:2010aa}. 
Furthermore, the cross-correlation will serve as an important 
 calibrator of systematics and biases in optical and infrared cosmic shear experiments 
 \cite{Vallinotto:2012aa, Das:2013ab}, which could otherwise limit future surveys \cite{Vallinotto:2013aa}.
 
 Measurements of CMB lensing have matured quickly in recent years. Its effects were 
 first detected in cross-correlation using radio-selected galaxy catalogs with \emph{WMAP} data 
 \cite{Hirata:2008aa, Smith:2007aa} and in auto-correlation using Atacama Cosmology 
 Telescope (ACT) data \cite{Das:2011ab}. Subsequent improvements to the lensing
 power spectrum were reported by the South Pole Telescope \cite{van-Engelen:2012aa}, ACT \cite{Das:2014aa},
  and the \emph{Planck} collaboration \cite{Planck-Collaboration:2014aa}.
 Further advances in CMB lensing data are expected from multiple experiments 
 in the near future \cite{Niemack:2010aa, Austermann:2012aa, Arnold:2010aa}. 
 Noting the anticipated enhancements of upcoming wide-field cosmic shear surveys \cite{DES, HSC}, 
 this work represents a first step in the application of a future, powerful tool for
 precision cosmology.

The measurement of the lensing-lensing cross power spectrum presented here
uses CMB data from the Atacama Cosmology Telescope and optical lensing data
 from the Canada-France-Hawaii Telescope (CFHT) Stripe 82 Survey (CS82). The 
 paper is structured as follows. Section \ref{sect:theory} presents a brief 
 overview of the theoretical expectation for the cross-correlation. 
 The lensing data used in this analysis are described in 
 section \ref{sect:maps}, and the analysis methods are detailed
 in section \ref{sect:methods}. The results of the measurement, as well as
 null tests and systematic checks, are outlined in section \ref{sect:results},
 and we conclude in section \ref{sect:conclusions}.

\section{Theoretical Background \label{sect:theory}}

The effects of cosmological gravitational lensing are encoded in the 
convergence field $\kappa$, which can be expressed as a weighted
projection of the matter overdensity $\delta$ \cite{Bartelmann:2001aa},

\begin{equation}
    \kappa(\mathrm{\bf{\hat{n}}}) = \int_0^{\infty} dz W^\kappa(z) \\
                \delta(\chi(z)\mathrm{\bf{\hat{n}}}, z).
\end{equation}
Assuming a flat universe, the lensing kernel $W^\kappa$ is

\begin{equation}
    W^\kappa(z) = \frac{3}{2}\Omega_{m} H_\circ^2 \\
            \frac{(1+z)}{H(z)} \frac{\chi(z)}{c} \int_z^{\infty} dz_s p_s(z_s) \\
                \frac{\chi(z_s) - \chi(z)}{\chi(z_s)},
\end{equation}
where $p_s(z)$ is the normalized redshift distribution of source galaxies, 
$\chi(z)$ is the comoving distance to redshift $z$, $\mathrm{\bf{\hat{n}}}$
is the direction on the sky, and $H_\circ$ and $\Omega_m$ are the 
present-day values of the Hubble and matter density parameters, respectively. 
We denote the kernel for the weak lensing of a source galaxy population with 
a redshift distribution $p_s(z) = dn/dz$ as $W^{\kgal}$.

For lensing of the CMB, the source redshift distribution can be 
approximated as $p_s(z) \simeq \delta_D(z-z_\star)$, where $z_\star \simeq
1090$ is the redshift of the surface of last scattering and $\delta_D$ is the
Dirac delta function. This yields the following kernel \cite{Lewis:2006aa}:

\begin{equation}
    W^{\kcmb}(z) =  \frac{3}{2}\Omega_{m}H_\circ^2  \frac{(1+z)}{H(z)} \frac{\chi(z)}{c} \\ 
    \left [ \frac{\chi(z_\star)-\chi(z)}{\chi(z_\star)} \right ].
    \label{eqn:cmb_kernel}
\end{equation}

\begin{figure}[t]
    \centering
        \includegraphics{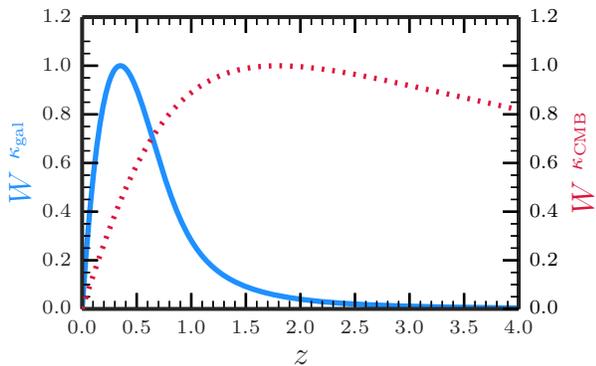}
    \caption{The lensing kernel $W^{\kgal}$ (solid) for the CS82 redshift distribution of source 
    galaxies (as given in Eq. \ref{nz}) and normalized to a unit maximum. For 
    comparison, the kernel for CMB lensing (Eq. \ref{eqn:cmb_kernel}) is shown as dashed, also normalized to 
    a unit maximum.}
    \label{fig:z_kernels}
\end{figure}

Using the Limber approximation \cite{Limber:1954aa, Kaiser:1992aa}, the 
cross power spectrum of the convergence fields due to CMB lensing and galaxy lensing 
can be computed to good precision as

\begin{equation}\label{eqn:theory_eqn}
    C_\ell^{\kcmb\kgal} = \int_0^\infty \frac{dz}{c} \frac{H(z)}{\chi(z)^2} W^{\kcmb}W^{\kgal} \\
                            P\left(k=\frac{\ell}{\chi}, z \right),
\end{equation}
where $P(k, z)$ is the matter power spectrum evaluated at wavenumber $k$ and
redshift $z$. The degree of cross-correlation between the two convergence
fields is determined by the overlap of the two kernels, weighted by the matter
power spectrum. For comparison, the CMB lensing kernel $W^{\kcmb}$ and the galaxy
lensing kernel $W^{\kgal}$ for the CS82 source population used in this work are 
shown in Fig. \ref{fig:z_kernels}. The mean redshift of the product
of $W^{\kgal}$ and $W^{\kcmb}$ is $z \sim 0.9$, illustrating that the cross 
power spectrum is sensitive to the amplitude of structure at 
intermediate redshifts.

\section{CMB and Galaxy Lensing Data \label{sect:maps}}

\subsection{ACT CMB Lensing Data}

ACT is a 6-meter telescope located in the Atacama desert in Chile  
\cite{Fowler:2007aa, Swetz:2011aa, Dunner:2013aa}. The CMB temperature maps used in this work 
are made from observations taken during 2008 - 2010 in the 148 GHz frequency channel 
and have been calibrated to 2\% accuracy as in \cite{Hajian:2011aa}. The maps are centered on 
the celestial equator with a width of 3 degrees in declination and 108 
degrees in right ascension and are identical to those used
in \cite{Das:2014aa}.

The lensing convergence fields are reconstructed from the CMB temperature maps
using the minimum variance quadratic estimator of \cite{Hu:2002ab} following the 
procedure used in \cite{Das:2011ab}. The lensing deflection induces correlations
in the Fourier modes of the previously uncorrelated, unlensed CMB. The lensing
convergence is estimated from these Fourier correlations with a quadratic 
estimator:

\begin{equation}\label{eqn:quad_estimator}
    \hat{\kappa}(\bm{L}) = N(\bm{L}) \int \mathrm{d}^2 \bm{l} \ f(\bm{L}, l) T(l)T(\bm{L}-\bm{l}),
\end{equation}
where $\bm{l}$ and $\bm{L}$ are Fourier space coordinates, $N$ is the normalization function, 
$T$ is the temperature field, and $f$ is a weighting function that maximizes the signal-to-noise ratio of the 
reconstructed convergence (see \cite{Hu:2002ab} for details). In the lensing reconstruction, 
we filter out temperature modes with a low signal-to-noise ratio, specifically those modes
below $\ell = 500$ and above $\ell = 4000$. This filtering does not prevent the
measurement of low-$\ell$ lensing modes, as the lensing signal at a given scale $\ell$ is obtained
from temperature modes separated by $\ell$ (see Eq. \ref{eqn:quad_estimator}). 
The maximum $\ell$ of included temperature modes is the only difference between
the lensing maps used in this work and those in \cite{Das:2014aa}.

The final normalization is obtained in a two step process, as in \cite{Das:2014aa}.
 A first-order approximation for the normalization is computed from the data power 
 spectrum, with an additional, small correction factor (of order 10\%) 
 applied from Monte Carlo simulations, which are designed to match both the
signal and noise properties of the ACT data. Finally, we obtain a simulated
mean field map $\langle \hat{\kappa} \rangle$ from 480 Monte Carlo realizations of 
reconstructed CMB lensing convergence maps and subtract this mean field from 
the reconstructed ACT lensing maps. The simulated mean field is non-zero due to 
noise and finite-map effects giving rise to a small ($\sim$5\%) artificial lensing signal, which must be subtracted.
Note that this set of 480 Monte Carlo realizations is also used to estimate error bars
on the final cross power spectrum measurement, as described in section \ref{sect:results}.

\subsection{CS82 Lensing Data}

\subsubsection{Data}

The Canada-France-Hawaii Telescope Stripe 82 Survey is an
$i'$-band survey of the so-called Stripe 82 region of sky along the
celestial equator \cite{Erben:2015-inpress-aa}.  The survey was
designed with the goal of covering a large fraction of Stripe 82 with
high quality $i'$-band imaging suitable for weak lensing
measurements. With this goal in mind, the CS82 survey was conducted
under excellent seeing conditions: the Point Spread Function (PSF) for
CS82 varies between $0.4\arcsec$ and $0.8\arcsec$ over the entire
survey with a median seeing of $0.6\arcsec$. In total, CS82 comprises
173 MegaCam $i'$-band images, with each image roughly one square
degree in area with a pixel size of 0.187 arcseconds. The area covered by the
survey is 160 degrees$^2$ (129.2 degrees$^2$ after masking out bright
stars and other image artifacts), and the completeness magnitude is
$i'\sim24.1$ (AB magnitude, 5$\sigma$ in a $2\arcsec$ aperture). Image
processing is largely based on the procedures presented in
\cite{Erben:2009aa, Erben:2013aa}. Weak lensing shear
catalogs were constructed using the state-of-the-art weak lensing
pipeline developed by the CFHTLenS collaboration which employs the
{\it lens}fit shape measurement algorithm \cite{Miller:2013aa,
  Heymans:2012aa}. We refer to these publications for more in-depth
details of the shear measurement pipeline.

\begin{figure*}[tb]
\includegraphics{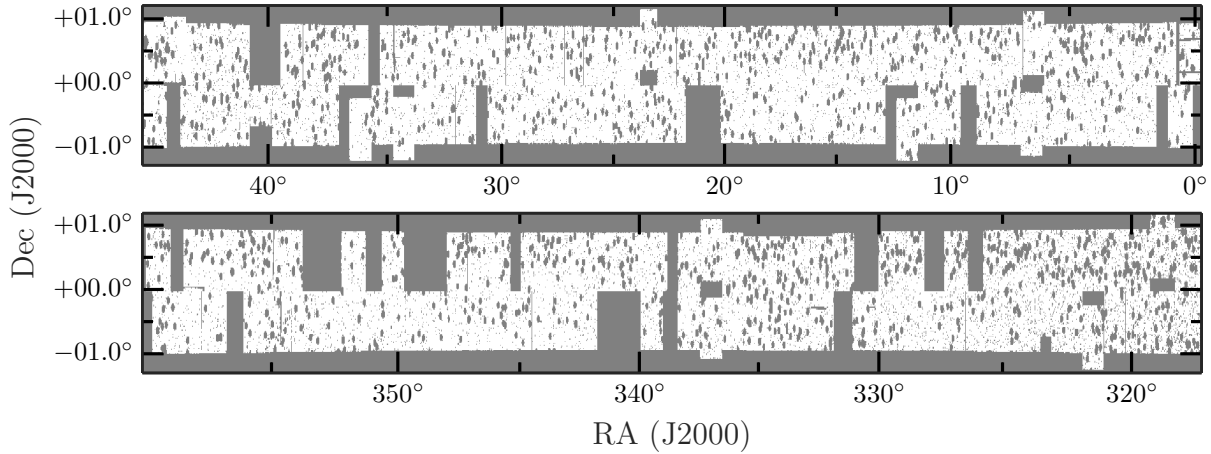}
\caption{The overlapping sky coverage of the ACT and CS82 data used in this work. Regions excluded as part of 
the CS82 mask are shown in grey, while unmasked regions, totaling 121 square degrees in area, are shown in white. }
\label{fig:sky_coverage}
\end{figure*}

Following \cite{Miller:2013aa} and \cite{Heymans:2012aa}, source
galaxies are selected to have $w>0$ and \texttt{FITSCLASS} $ = 0$. Here, $w$
represents an inverse variance weight accorded to each source galaxy
by {\it lens}fit, and \texttt{FITSCLASS} is a flag to both remove stars and
select galaxies with well-measured shapes (see details in
\cite{Miller:2013aa}). After these cuts, the CS82 source galaxy
density is 15.8 galaxies arcmin$^{-2}$, and the effective weighted
galaxy number density (see equation 1 in \cite{Heymans:2012aa}) is
12.3 galaxies arcmin$^{-2}$. Note that these numbers do not include
any cuts on photometric redshift quality since for the purposes of
this paper, we only need to know the CS82 source galaxy redshift distribution
(see the following section). We derive the multiplicative shear calibration
factor $m$ in the same manner as \cite{Miller:2013aa}, with the
multiplicative shear measurement bias equal to $1+m$.

Our reduction pipeline includes an automated masking routine to detect
artifacts on an image-by-image basis and to mask out bright stars
\cite{Erben:2013aa}. Each mask is manually inspected and modified
when necessary (for example, to mask out faint satellite trails) to
create a final set of masks. These high-resolution masks are then
re-binned to a resolution of 1 arcminute and combined into a larger
single mosaic mask map for the full CS82 data. This mask is shown 
in Figure \ref{fig:sky_coverage}, which shows the overlapping  
sky coverage of the ACT and CS82 data used in this work. The
total area of the overlapping, unmasked region is 121 square degrees.

\subsubsection{Source redshift distribution \label{sect:dndz}}

\begin{figure*}[tb]
\includegraphics[scale=0.5]{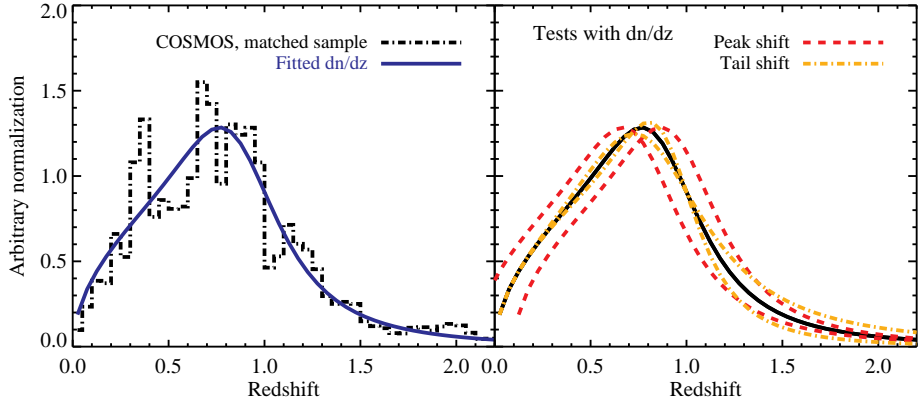}
\caption{Redshift distribution of CS82 source galaxies. Left: redshift
  distribution for a matched sample of galaxies from the COSMOS
  survey. The blue solid line indicates our fit to the COSMOS matched
  sample.  Right: we test how the amplitude of the theoretical
  lensing-lensing cross power spectrum changes when we vary the peak of
  the $dn/dz$ (red, dashed lines) and the high-redshift tail of the
  distribution (orange, dash-dotted lines). These variations in the $dn/dz$ lead to
  changes of order $10-20\%$ in the amplitude of the theoretical
  lensing-lensing cross power spectrum. Accurate estimates of $dn/dz$
  will be crucial for this kind of cross-correlation in the future.}
\label{cs82dndz}
\end{figure*}

As the CS82 $i'$-band imaging is deeper than the overlapping
multi-color co-add data from SDSS \cite{Annis:2014aa}, we cannot
estimate a photometric redshift for each galaxy in our source
catalog. However, for the purposes of this work, we do not require a
photometric redshift estimate for each source galaxy. Instead, only
the source redshift distribution is needed to predict the amplitude of
the cross-correlation. We estimate this redshift distribution using
the 30-band COSMOS photometric redshift catalog
\cite{Ilbert:2009aa}. We select a random sample of COSMOS galaxies
such that the $i'$-band magnitude distribution of the random sample matches
the CS82 source catalog. We then fit the $dn/dz$ from this matched sample,
weighting each galaxy by $w$, the inverse variance weight accorded to
each CS82 source galaxy. By using this weight, we account for the
increase in the shape measurement noise at faint magnitudes (see
equation 8 in \cite{Miller:2013aa}).  Adopting the functional form
from \cite{Fu:2008aa}, the weighted source redshift distribution is
given by:

\begin{equation}\label{nz}
\frac{dn}{dz} = A \frac{z^a + z^{ab}}{z^b + c},
\end{equation}
with $a = 0.531$, $b = 7.810$, $c = 0.517$, and $A = 0.688$. The
source redshift distribution from the matched COSMOS sample is shown
in Fig. \ref{cs82dndz}.


There are uncertainties in our $dn/dz$ estimate due to sample variance
in the COSMOS data, errors in the COSMOS photometric redshifts, and
the assumed parametric form for $dn/dz$. Estimating these
uncertainties is a nontrivial task and is beyond the scope of this
paper, as the main goal of this work is simply to present the
detection of the cross-correlation. Nonetheless, to give some sense of
the effects of uncertainty in $dn/dz$, we investigate how the predicted amplitude of
the cross-correlation varies when we shift the peak and the high-redshift tail of $dn/dz$. 
For these tests, we shift the peak of $dn/dz$
by $\Delta z = \pm 0.1$ and shift the high-redshift tail of $dn/dz$ by
varying the parameter $b$ by $\pm$30\%. These four test cases are shown in the right
panel of Fig. \ref{cs82dndz}. Again, we stress that these tests are
not necessarily designed to represent the true underlying uncertainty
in our $dn/dz$ estimate (which is nontrivial to compute), but only to give some
idea of how variations in $dn/dz$ can affect the predicted amplitude
of the cross-correlation.

When computing the theoretical cross power spectrum with fixed
cosmological parameters using Eq. \ref{eqn:theory_eqn}, we find that
these $dn/dz$ variations lead to changes of order $10-20\%$ in the
amplitude of the theory curve. The largest amplitude change occurs
when shifting the tail of the source distribution to higher redshift,
with the other variations leading to comparable changes. As the CMB
lensing kernel $W^{\kcmb}$ peaks at $z \sim 2$ with a broad tail to
higher redshift, the degree of cross-correlation is quite sensitive to
the tail of the source galaxy redshift distribution. Clearly, the
interpretation of our results depends on the assumed $dn/dz$.  In
general, the high-redshift tail of the source redshift distribution is
notoriously difficult to measure from photometric surveys. This is due
in part to the Lyman-Balmer break degeneracy in photometric redshift codes for
galaxies at $z \gtrsim 1.5$ (which requires difficult to obtain deep
near-infrared or $U$-band imaging to be resolved), but also because
high-redshift galaxies are faint and thus have more unreliable
photometric redshifts. In conclusion, it is clear that future measurements of this
kind will need to pay particular attention to systematics associated
with the source redshift distribution.

\subsubsection{CS82 shear maps}

We create a series of maps for the CS82 data that follow a regular grid with a
pixel size of 1 arcminute and that are matched to the mosaic mask map
described previously. To create shear maps, we closely follow the
procedure outlined in \cite{Van-Waerbeke:2013aa} to account for the
multiplicative shear measurement bias ($1+m$) and the weighting $w$. A
normalized ellipticity map $M_\mathrm{e1}$ is constructed for the
$e_1$ component of the ellipticity by summing $e_1$ over all source
galaxies within in each pixel $(x,y)$ and normalizing as 

\begin{equation}
    \label{eqn:normed_shear}
M_\mathrm{e1}(x,y) = \frac{\sum_i w_i e_{1,i}}{\sum_i w_i(1+m_i)},
\end{equation}
where $w_i$ is the inverse variance weight and $m_i$ is the shear 
calibration factor associated with the $i^\mathrm{th}$ galaxy \cite{Miller:2013aa}. 
In a similar fashion, we also compute the following maps:

\begin{itemize}
\item $M_{\rm e2}(x,y)$: similar to $M_{\rm e1}(x,y)$ but for the
  $e_2$ component of the ellipticity.
\item $M_{\rm psf1}(x,y)$: similar to $M_{\rm e1}(x,y)$ but $e_1$ is replaced
  by an estimate of the $e_1$ component of the PSF ellipticity at
  galaxy position $i$.
\item $M_{\rm psf2}(x,y)$: similar to $M_{\rm e2}(x,y)$ but $e_2$ is replaced
  by an estimate of the $e_2$ component of the PSF ellipticity at
  galaxy position $i$.
\item $M_{\rm bmode1}(x,y)$: the first B-mode component of the ellipticity, which
is equal to $-M_{\rm e2}(x,y)$.
\item $M_{\rm bmode2}(x,y)$: the second B-mode component of the ellipticity, which
is equal to $M_{\rm e1}(x,y)$.
\end{itemize}

We also create a set of 500 random maps for each component of the
ellipticity. In these maps, the position of each
galaxy is preserved, but for each realization, we rotate source
galaxies by a random position angle. This process ensures that the maps have the same
shape noise as the CS82 ellipticity maps but do not contain a
cosmological shear signal. We use these random maps for null tests in
cross-correlation with the true ACT data (see section
\ref{sect:results}).

Finally, as discussed in \cite{Heymans:2012aa}, 25\% of the CFHT Legacy
Survey fields have a significant PSF residual and are rejected for cosmic
shear studies \cite{Heymans:2013aa, Benjamin:2013aa}. However, this
cross-correlation analysis should be much less sensitive to
PSF-related errors in comparison to cosmic shear measurements because
CS82 PSF patterns should be uncorrelated with the ACT CMB lensing
signal. In addition, \cite{Van-Waerbeke:2013aa} found that it was not
necessary to reject these fields. Nonetheless, it is possible that the
PSF pattern correlates with the ACT signal simply by chance. The
$M_{\rm psf}(x,y)$ maps are designed to test and rule out this
possibility (see the discussion of null tests in section
\ref{sect:results}).

\section{Methods \label{sect:methods}}

\subsection{Power Spectrum Estimation}

The cross-correlation of the ACT CMB lensing and CS82 galaxy lensing
convergence fields is computed in Fourier space. The choice to
reconstruct the correlation in Fourier space rather than real space
was made in order to limit correlations between different data bins,
which can complicate the interpretation of the final
measurement. Furthermore, this method minimizes the total number of
Fourier transforms needed, which reduces noise due to windowing and
edge effects. In order to obtain an unbiased estimate of the cross spectrum, we
follow a procedure similar to the steps outlined in previous ACT power
spectrum analyses \cite{Das:2014aa, Das:2011aa}, which properly
account for the coupling of Fourier modes induced by filtering and
windowing effects. The notation and terminology in this section closely 
follows that of these previous ACT analyses. 

\subsubsection{The Data Window}\label{datawindow}

First, the real space ACT convergence map is repixelized to match the
resolution (1 arcminute) of the CS82 data. 
Then, the ACT data and CS82 ellipticity maps are
spatially divided into two noncontiguous patches on which the cross
spectrum estimation is computed separately. The two patches are
divided at zero right ascension due to a coincidental discontinuity in the CS82 
imaging at this location. This divides the original map into two roughly equal area
patches. We denote the separate patches with greek indices, such
that patch $\alpha$ of the two CS82 ellipticity maps at position
$\bm{\theta} = (x, y)$ is denoted as $M^\alpha_{e1}(x, y)$ and
$M^\alpha_{e2}(x, y)$. Similarly, patch $\alpha$ of the repixelized
ACT convergence map is denoted as $M^\alpha_\mathrm{\kcmb}(x,y)$.

Both the CS82 and ACT data patches are multiplied in real space by a tapering 
function and the CS82 mask map, which masks out image artifacts and bright
point sources. The tapering function minimizes noise introduced by the patch
edges in Fourier space. It is generated by convolving a map that is
unity in the center and zero over 10 pixels at the edges with a
Gaussian of full width at half maximum of $5^\prime$. 
The window function in real space is the product of these 
two components -- the tapering function and the CS82 mask. 
 In the following discussion, the window function 
is denoted by $K^\alpha$ and the windowed data patches are 
denoted as $\widetilde{M}^\alpha_{i}$, where $i \in [e1, e2, \kcmb]$.

\subsubsection{Galaxy Lensing Convergence Reconstruction}

We reconstruct the CS82 convergence field in Fourier space from the windowed ellipticity 
patches, following the prescription outlined in \cite{Kaiser:1993aa}. 
The galaxy lensing convergence field in Fourier space $\widetilde{M}^\alpha_{\kgal}(\bm{\ell})$
is given by

\begin{equation}\label{eqn:kaiser_squires}
   \widetilde{M}^\alpha_{\kgal}(\bm{\ell}) = F_\ell \left [ \widetilde{M}^\alpha_{e1}(\bm{\ell}) \\ 
                                        \frac{\ell_x^2 - \ell_y^2}{\ell^2} + \\
                                        \widetilde{M}^\alpha_{e2}(\bm{\ell}) \\ 
                                         \frac{2\ell_x\ell_y}{\ell^2}  \right ], 
\end{equation}
where the wavevector $\bm{\ell} = (\ell_x, \ell_y) =  2\pi/\bm{\theta}$ is defined as the 
two-dimensional Fourier analog of $\bm{\theta}$, $\ell^2 = \ell_x^2 + \ell_y^2$, and 
$F_\ell$ is a Gaussian smoothing filter of full width at half maximum of $2^\prime$.

\subsubsection{Mode-coupling}

A 2D pseudo-spectrum is computed from the windowed convergence fields as

\begin{equation}
    \widetilde{C}^{\kcmb\kgal}_\ell = \mathrm{Re}\left [ \widetilde{M}^\star_{\kcmb}(\bm{\ell}) \\
                                                \widetilde{M}_{\kgal}(\bm{\ell}) \right],
\end{equation}
where the patch index has been suppressed for clarity. The 1D binned spectrum
$\widetilde{C}_b$ is computed by averaging the 2D spectrum in annular bins

\begin{equation}\label{eqn:binning}
    \widetilde{C}_b^{\kcmb\kgal} = \sum_{\bm{\ell}} P_{b \ell} \widetilde{C}^{\kcmb\kgal}_\ell,
\end{equation}
where $P_{b \ell}$ is the binning matrix, which is defined to be zero when $\bm{\ell}$
is outside the annulus defined by bin index $b$ and unity otherwise.

Noting that the windowing operation in real space
corresponds to a convolution in Fourier space and using Eqs. \ref{eqn:kaiser_squires} 
to \ref{eqn:binning}, we can express the binned 
1D pseudo-spectrum $\widetilde{C}_b$ in terms of the underlying spectrum $C_\ell$ as 
\begin{equation}
   \widetilde{C}_b^{\kcmb\kgal} = \sum_{\bm{\ell}, \bm{\ell^\prime}} P_{b\ell} \\
                                |K(\bm{\ell} - \bm{\ell}^\prime)|^2 F_{\ell^\prime} C_{\ell^\prime}^{\kcmb\kgal},
\end{equation}
where $K$ is the three-component window function discussed previously. 
We relate this quantity to a binned version of the true spectrum $C_b$ via an inverse binning
operator $Q_{\ell b}$, which is unity when $\ell \in b$ and zero otherwise, 

\begin{eqnarray}
    \widetilde{C}_b^{\kcmb\kgal} & = & \sum_{\bm{\ell}, \bm{\ell^\prime}, b^\prime} P_{b\ell}
                                 |K(\bm{\ell} - \bm{\ell}^\prime)|^2 F_{\ell^\prime} Q_{\ell^\prime b^\prime}
                                  C_{b^\prime}^{\kcmb\kgal}, \nonumber \\ 
                                 & \equiv & \sum_{b^\prime} M_{bb^\prime}C_{b^\prime}^{\kcmb\kgal},
\end{eqnarray}
where $M_{b b^\prime}$ is the mode-coupling matrix, which is well-behaved and stable to 
inversion. Finally, we define the unbiased estimator of the power
spectrum (denoted by a circumflex) as 

\begin{equation}\label{eqn:final_power_spectrum}
    \widehat{C}_b^{\kcmb\kgal} = \sum_{b^\prime} M_{bb^\prime}^{-1} \widetilde{C}_{b^\prime}^{\kcmb\kgal}.
\end{equation}
We use Eq. \ref{eqn:final_power_spectrum} to estimate the cross power spectrum
for each patch and compute the final cross power spectrum as the mean of the spectra 
from the two individual patches.

\subsection{Pipeline Validation}

\begin{figure}[t]
    \centering
        \includegraphics[scale=1]{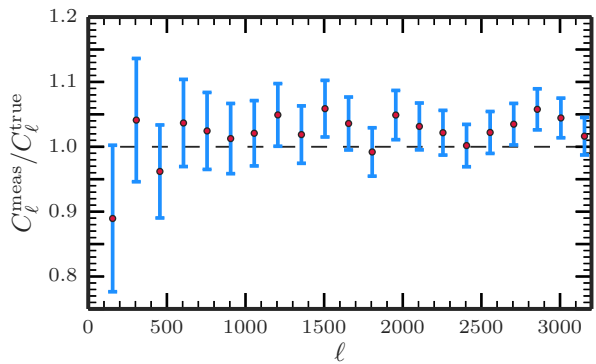}
    \caption{The ratio of the auto power spectrum of simulated lensing
    convergence maps measured using our analysis pipeline $C_\ell^\mathrm{meas}$ 
    to the true, input auto spectrum $C_\ell^\mathrm{true}$. The analysis pipeline described in
    section \ref{sect:methods} properly recovers the input spectrum from simulations.}
    \label{fig:sim_pipeline_validation}
\end{figure}

We use simulated galaxy lensing maps to validate the power spectrum analysis 
steps described in the previous section. The simulated maps are constructed using the shear 
signal from the N-body simulations described in \cite{Harnois-Deraps:2012aa}. 
Projected shear and convergence ``tiles'' are produced for 25 separate lines of sight 
in the simulation. Each tile covers an area of 12.8 $\mathrm{deg^2}$ 
and has a pixel size of $0.21^\prime$. For simplicity, we use shear and convergence maps
constructed using source galaxies at a single redshift of $z=0.73$. 
As the purpose of the simulation maps is only
to verify the analysis pipeline, a more realistic $dn/dz$ is not required. 

We repixelize the 25 simulated tiles to match the pixel size of the
CS82 data (1 arcminute) and use these tiles to construct a map
with equal size and area to the map used in the data analysis. We then
multiply the shear maps by the CS82 mask map, reconstruct 
the convergence in Fourier space using Eq. \ref{eqn:kaiser_squires}, and 
estimate the convergence auto power spectrum using the analysis
steps outlined in the previous section. Fig.
\ref{fig:sim_pipeline_validation} shows the result of this
calculation, plotting the ratio of the reconstructed
 auto spectrum to the true, input spectrum.  
The analysis pipeline accurately recovers the input power 
spectrum, within the measured errors.

\section{Results \label{sect:results}}

\subsection{The CMB Lensing - Galaxy Lensing Cross Power Spectrum}

\begin{figure}[tb]
      \centering
          \includegraphics[scale=1]{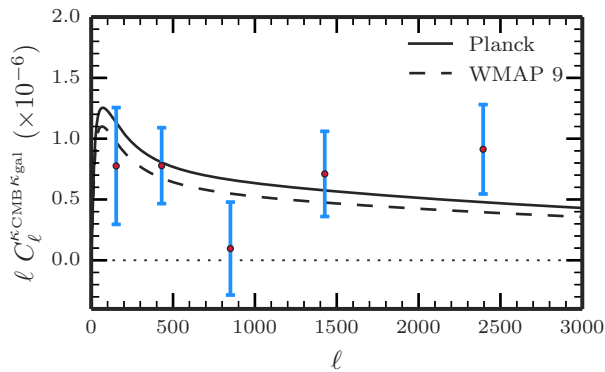}
          \caption{The CMB lensing - galaxy lensing convergence cross
            power spectrum (red points), measured using ACT and CS82
            data. Error bars are computed using Monte Carlo methods (see text), and
            the significance of the measurement is $4.2\sigma$. 
            The solid and dashed black lines show the
            expected power spectra assuming the \emph{Planck} + 
            lensing + WP + highL and \emph{WMAP}9 + eCMB cosmological models, respectively. 
            The theoretical spectra shown correspond to $A = 1$, and relative
            to these models, the best-fit amplitudes to our data
            are $A^\mathrm{Planck} = 0.78 \ \pm \ 0.18$ and $A^\mathrm{WMAP} = 0.92\ \pm \ 0.22$.}
      \label{fig:cross_power_spectrum}
  \end{figure}

The cross power spectrum of the ACT CMB lensing and CS82 galaxy
lensing convergence maps is shown in Fig. \ref{fig:cross_power_spectrum}. 
The error bars on the data points are computed by cross-correlating 
480 Monte Carlo realizations of simulated reconstructed CMB lensing maps with the 
true CS82 data. The simulated CMB lensing maps are constructed to match 
both the signal and noise properties of the ACT data maps. As a consistency check, 
we note that these error bars agree well with the theoretical expectation, 
which is computed using the auto spectra of the individual maps. Specifically, 
the analytic error bars are proportional to $\sqrt{C_\ell^{\kcmb\kcmb}C_\ell^{\kgal\kgal}}$ 
and an additional factor that accounts for the number of independent pixels in each data bin. 
We assume in both methods that the maps are uncorrelated, which is a valid approximation
since both maps are noisy such that $C_\ell^{\kcmb\kcmb}C_\ell^{\kgal\kgal} \gg 
(C_\ell^{\kcmb\kgal})^2$. Using Monte Carlo methods to estimate the error bars allows 
us to calculate the full covariance matrix. Neighboring bins are approximately 
$\sim$10\% anti-correlated,  while nearly all other off-diagonal correlations are 
less than $5\%$ of the bin auto-correlation. We account for the full covariance 
matrix when computing measurement significances.

The theoretical expectation for the cross power spectrum obtained by evaluating
Eq. \ref{eqn:theory_eqn} is also shown in Fig. \ref{fig:cross_power_spectrum}.
We consider two separate cosmological models for comparison. 
First, we use the best-fit \emph{Planck} + lensing + WP + highL parameter set 
with $\sigma_8 = 0.827$ \cite{Planck-Collaboration:2014ac}, where WP refers to 
the inclusion of \emph{WMAP} polarization data and highL refers to the inclusion
of ACT and South Pole Telescope high-$\ell$ CMB data in the parameter likelihood. 
Second, we consider the \emph{WMAP}9 + extended CMB (eCMB) model with
$\sigma_8 = 0.81$ \cite{Hinshaw:2013aa}, where eCMB refers to the usage 
of high-$\ell$ ACT and South Pole Telescope CMB data in the parameter likelihood. 
In both calculations, the non-linear matter power spectrum 
(HALOFIT, \cite{Smith:2003aa, Takahashi:2012aa}) is used. 

We define a parameter $A$ for the amplitude of the cross spectrum relative
to the two models considered here, defined such that $A = 1$ 
corresponds to the fiducial model.
We compute the amplitude likelihood for both the \emph{Planck} and \emph{WMAP} models, 
assuming no uncertainties in the CS82 source distribution.  
Relative to the \emph{Planck} fiducial model, we obtain a best-fit amplitude 
$A^\mathrm{Planck} = 0.78 \ \pm \ 0.18$, with $\chi^2 = 3.58$ and
$\chi^2/\nu = 0.90$ for $\nu = 4$ degrees of freedom. 
Relative to the \emph{WMAP}9 model, we measure an amplitude 
$A^\mathrm{WMAP} = 0.92 \ \pm \ 0.22$, with $\chi^2 = 3.68$ and
$\chi^2/\nu = 0.92$. The significance is computed as the square-root of the
difference between the chi-squared values of the null line ($A = 0$) and the 
best-fit theoretical spectrum: 
$(\Delta \chi^2)^{1/2} = \sqrt{\chi^2_\mathrm{null} - \chi^2_\mathrm{theory}}$.
With a measured value of $\chi^2_\mathrm{null} = 21.67$, the best-fit 
theoretical model is favored over the null hypothesis with a significance of 
$4.2\sigma$ (for both the \emph{Planck} and \emph{WMAP}9 models). 

 
 Since the amplitude of the cross spectrum scales as the square of the amplitude 
 of density fluctuations, this measurement corresponds to a 
  $\sim$12\% constraint on the amplitude of structure at intermediate redshifts, 
  $z \sim 0.9$, which corresponds to the mean redshift of the product of 
  the CMB lensing and 
  galaxy lensing kernels (see Fig. \ref{fig:z_kernels}). This constraint 
  on $\sigma_8$ is given as an approximate benchmark for comparison to other 
  current growth of structure measurements, rather than a robust cosmological 
  constraint. We choose not to perform a 
  more detailed cosmological interpretation of this measurement for several reasons.
  As discussed in section \ref{sect:maps}, uncertainties in the source 
  galaxy redshift distribution must be well-understood before achieving
  accurate constraints, as $dn/dz$ errors propagate to uncertainties in the predicted amplitude 
  of the theoretical cross power spectrum.  Furthermore, as noted recently in 
  \cite{Hall:2014aa, Troxel:2014aa}, the CMB lensing -- galaxy lensing power spectrum
  is contaminated, at some level, by a cross-correlation term between CMB lensing and 
  galaxy intrinsic alignment. The magnitude 
  of this contamination must be carefully calibrated before using the CMB lensing -- galaxy 
  lensing cross-correlation for precision cosmology in the future.

\subsection{Null Tests}

\begin{figure}[tb]
    \centering
        \includegraphics[scale=1]{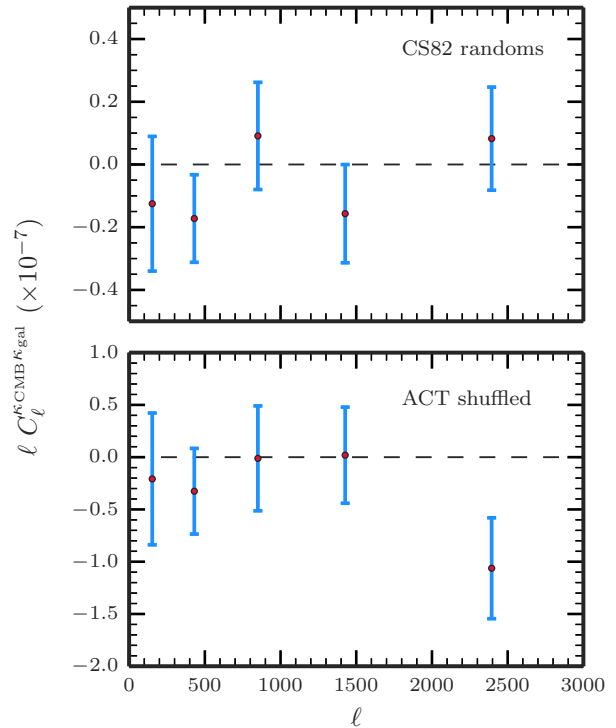}
    \caption{Two successful null tests, both consistent with zero. Top: the 
    mean correlation between 500 randomized galaxy lensing maps  and the true
    ACT data. Bottom: the mean correlation between the true CS82 data
    and 58 ACT ``shuffled'' maps, constructed by shifting the data in
    intervals of $0.75^\circ$ along the right ascension direction. The 
    probabilities to exceed the measured $\chi^2$ for these tests are 
    63\% and 34\%, respectively. Note that the scaling of the 
    $y$-axis here is an order of magnitude smaller than the $y$-axis 
    of Fig. \ref{fig:cross_power_spectrum}. }
    \label{fig:random_nulls}
\end{figure}

\begin{figure}[tb]
    \centering
        \includegraphics[scale=1]{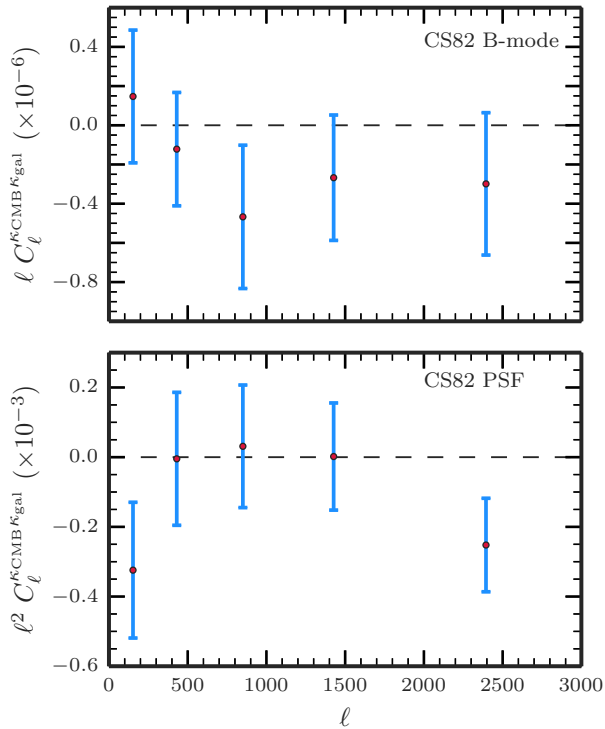}
    \caption{Two tests of the CS82 galaxy lensing data, both consistent with zero. 
    The data points show the correlations between the ACT data and the 
    CS82 B-mode ellipticity data (top) and the PSF ellipticity data (bottom).  
     The probabilities to exceed the measured $\chi^2$ for these tests are 55\% and 13\%, respectively. 
     Note that the quantity $\ell^2C_\ell$ is plotted in the bottom panel in order 
     to increase the dynamic range of the plot.}
    \label{fig:cs82_nulls}
\end{figure}

We verify our pipeline and measured cross power spectrum with a series of null tests. 
The first test uses 500 realizations of randomized galaxy lensing shear maps,
described previously in section \ref{sect:maps}. We compute the mean cross power spectrum
 between the true ACT convergence field and these random maps.
 Shown in the top panel of Fig. \ref{fig:random_nulls},
this mean correlation is consistent with zero, with $\chi^2 = 3.4$ for five
degrees of freedom; the probability of random deviates with the same 
covariances to exceed this chi-squared is 63\%. Note that the set of 500 randomized
shear maps do not contain a cosmological shear signal and thus, can only be used 
as a null test rather than to estimate error bars for the final cross spectrum measurement. 
We also create a set of 58 ``shuffled'' ACT maps by shifting the true ACT data 
in intervals of $0.75^\circ$ along the right ascension direction. The mean of 
the cross-correlation between these shuffled maps and the CS82 convergence data 
is shown in the lower panel of Fig. \ref{fig:random_nulls}. This mean correlation is 
also consistent with null signal, with $\chi^2 = 5.7$ and a probability to exceed 
of 34\%. The error bars for each of these measurements are computed
using the full covariance matrix as determined from the Monte Carlo realizations, as 
was done for the true data.

We also perform two specific tests of the CS82 shear data, designed
to check for any unexpected correlations due to possible systematic issues
with the galaxy lensing data. We compute the cross
power spectrum using the same methods outlined in section \ref{sect:methods}, but replace
the CS82 ellipticity data with 1) the B-mode ellipticity maps $M_\mathrm{bmode1/2}$
and 2) the PSF ellipticity maps $M_\mathrm{psf1/2}$. The B-mode ellipticity is 
obtained using the transformation $(e_1, e_2)$ to $(-e_2, e_1)$, and in the absence
of systematics, should vanish. The cross power spectrum between the ACT data
and the B-mode convergence data is shown in the top panel of Fig. \ref{fig:cs82_nulls}.
As expected, the measurement is consistent with null signal, with $\chi^2 = 4.0$ for 
five degrees of freedom, corresponding to a 55\% probability that the chi-squared
of random noise would exceed the measured value.
The bottom panel of Fig. \ref{fig:cs82_nulls} shows the correlation 
between the ACT data and the PSF data. The result is also consistent
with zero, with $\chi^2 = 8.6$ and a probability to exceed of 13\%. The error 
bars for both spectra are computed from Monte Carlo estimates, as done previously.

\section{Conclusions \label{sect:conclusions}}

We have cross-correlated CMB lensing and galaxy lensing convergence maps and 
measured the lensing-lensing cross power spectrum for the first time -- at $4.2\sigma$ 
significance. This cross-power is a direct gravitational measurement of the distribution of mass 
at redshifts $\sim$ 0.9. The measurement constrains the amplitude of structure to an 
uncertainty of $\sim$12\%, although contamination from galaxy intrinsic alignments
and errors in the $dn/dz$ must be carefully considered for more precise cosmological constraints. 
Our method is remarkably robust to instrumental and astrophysical systematic errors.
It is performed with a cross-correlation of mass measurements relying on completely 
different measurement techniques and photon wavelengths, which few systematics can survive. 
Despite the moderate detection significance, this robustness makes a first measurement of 
this cross-correlation signal a valuable confirmation of the $\Lambda$CDM model for large-scale 
structure at intermediate redshifts.

In just the next few years, measurements of lensing-lensing cross-correlations 
are expected to increase in signal-to-noise by more than an order of magnitude 
\cite{Niemack:2010aa, Austermann:2012aa, Arnold:2010aa, DES, HSC}. CMB lensing-galaxy lensing 
cross-correlations have the potential to greatly contribute to cosmology in two main ways. 
First, they can serve as a calibrator of instrumental systematics, which 
may potentially limit future optical and infrared weak lensing surveys. 
By adding information from lensing-lensing cross-correlations to weak lensing 
power spectra, additive and multiplicative biases can be precisely constrained, 
which will allow future weak lensing surveys to reach their full cosmological 
potential (see e.g., \cite{Vallinotto:2012aa, Das:2013ab}). Second, they will serve as an independent, 
robust measurement of the amplitude of structure at intermediate redshifts. 
When combined with probes at higher redshift (e.g., CMB lensing) and 
lower redshift (e.g, weak lensing), lensing-lensing    
cross-correlations will help measure the growth of structure across a   
wide range of redshifts. This, in turn, will allow for powerful constraints 
on the sum of neutrino masses and the properties of dark energy. This 
work thus demonstrates an important proof of concept of an exciting new 
cosmological probe.
\begin{acknowledgments}
We thank the CFHTLenS team for their pipeline development and
verification upon which much of this survey pipeline was built. We also thank 
Jeff Newman and Peter Freeman for helpful conversations about statistical analysis. 
This work was supported by the U.S. National Science Foundation through
awards AST-0408698 and AST-0965625 for the ACT project, as well as
awards PHY-0855887 and PHY-1214379.  Funding was also provided by
Princeton University, the University of Pennsylvania, and a Canada
Foundation for Innovation (CFI) award to UBC. ACT operates in the
Parque Astron\'omico Atacama in northern Chile under the auspices of
the Comisi\'on Nacional de Investigaci\'on Cient\'ifica y
Tecnol\'ogica de Chile (CONICYT). Computations were performed on the
GPC supercomputer at the SciNet HPC Consortium. SciNet is funded by
the CFI under the auspices of Compute Canada, the Government of
Ontario, the Ontario Research Fund -- Research Excellence; and the
University of Toronto. This work was based on observations obtained
with MegaPrime/MegaCam, a joint project of CFHT and CEA/DAPNIA, at the
Canada-France-Hawaii Telescope (CFHT), which is operated by the
National Research Council (NRC) of Canada, the Institut National des
Science de l'Univers of the Centre National de la Recherche
Scientifique (CNRS) of France, and the University of Hawaii. The
Brazilian partnership on CFHT is managed by the Laborat\'{o}rio Nacional
de Astrof\`isica (LNA). This work made use of the CHE cluster, managed
and funded by ICRA/CBPF/MCTI, with financial support from FINEP and
FAPERJ. We thank the support of the Laborat\'{o}rio Interinstitucional
de e-Astronomia (LIneA). NH is supported by the National Science Foundation Graduate 
Research Fellowship under grant number DGE-1106400 and the Berkeley Fellowship for Graduate Study. 
TE is supported by the Deutsche Forschungsgemeinschaft through project ER 327/3-1 and by the
Transregional Collaborative Research Centre TR 33 – ``The Dark
Universe''. BM acknowledges financial support from the CAPES Foundation grant 12174-13-0.
CH acknowledges support from the European Research Council
under the EC FP7 grant number 240185. HYS acknowledges the support from 
Marie-Curie International Incoming Fellowship (FP7-PEOPLE-2012-IIF/327561), Swiss
National Science Foundation (SNSF) and NSFC of China under 
grants 11103011. This work was supported by World Premier International 
Research Center Initiative (WPI Initiative), MEXT, Japan. HH is supported by the Marie Curie IOF 252760, by a CITA
National Fellowship, and the DFG grant Hi 1495/2-1. JPK acknowledges
support from the ERC advanced grant LIDA and from CNRS. We acknowledge support
from NSF Grant 1066293 and thank the Aspen Center for Physics for hospitality
during the writing of this paper.
\end{acknowledgments}

\bibliography{}

\end{document}